\documentclass{aa}
\usepackage{epsfig}
\usepackage{amssymb}

\newcommand{\Pm}{P_m}
\newcommand{\Rey}{\mbox{\it Re}}
\newcommand{\Rm}{\Rey_m}
\newcommand{\bnabla}{\mbox{\boldmath $\nabla$}}
\newcommand{\Eccf}{E_{\mbox{\it\scriptsize CCF}}}
                                        
\newcommand{\beq}{\begin{equation}}
\newcommand{\eeq}{\end{equation}}
\newcommand{\beqarr}{\begin{eqnarray}}
\newcommand{\eeqarr}{\end{eqnarray}}
\newcommand{\barr}{\begin{array}}
\newcommand{\earr}{\end{array}}
\newcommand{\bcent}{\begin{center}}
\newcommand{\ecent}{\end{center}}
\newcommand{\rf}[1]{(\ref{#1})}

\newcommand{\vechat}[1]{{\skew3\hat{\vec{#1}}}}

\newcommand{\laplace}{\nabla^2}  

\newcommand{\cross}{\wedge}
\newcommand{\grad}{\bnabla}
\newcommand{\dvgnce}{\bnabla \cdot}
\newcommand{\curl}{\bnabla \wedge}

\newcommand{\pder}[3]{\frac{\partial^{#3}#1}{\partial #2^{#3}}}
\newcommand{\pd}[1]{\partial_{#1}}

\newcommand{\bess}{{\mathcal{B}}}
\newcommand{\ex}{{\mathrm e}}
\newcommand{\im}{{\mathrm i}}

\newcommand{\grB}{\sigma\hspace{-1.5pt}_B\hspace{1.2pt}}

\title{A Taylor-Couette Dynamo}
\author{A. P. Willis \and C. F. Barenghi}
\institute{Department of Mathematics,  
 University of Newcastle, Newcastle upon Tyne, NE2 7RU, UK}

\date{Received 19 April 2002 / Accepted 8 July 2002}

\begin{document}

\abstract{
   Recent experiments have shown that it is possible to study
   a fundamental astrophysical process such as dynamo action in
   controlled  laboratory conditions using simple MHD flows.
   In this paper we explore the possibility that Taylor-Couette flow,
   already proposed as a model of the magneto-rotational instability of
   accretion discs, can sustain generation of magnetic field. Firstly, by
   solving the kinematic dynamo problem, we identify the region of
   parameter space where the magnetic field's growth rate is higher.
   Secondly, by solving simultaneously the coupled nonlinear equations
   which govern velocity field and magnetic field, we find a
   self-consistent nonlinearly saturated dynamo.
}

\maketitle

\section {Motivation}

Even though astrophysical objects possessing observable magnetic fields
are extremely diverse, it is widely accepted that the physical 
mechanisms supporting the fields are fairly universal and rely on
features common to virtually all astrophysical objects (e.g.\ differential
rotation, convective or turbulent motions, etc.).
MHD dynamo theory quantifies this idea and states that astrophysical
magnetic fields are created by inductive currents driven by motions of
electrically conducting plasmas (Moffatt 1978).
Until recently, dynamo action was the subject of theoretical or
numerical investigations only.  The recent demonstration of dynamo
action in controlled laboratory experiments
(Gailitis et al. 2001; 
 Stieglitz \& Muller 2001)
has stimulated interest in the study of dynamo action in confined
geometries of potential laboratory interest, such as spheres or 
cylinders.  The configuration of an MHD fluid confined between
concentric cylinders (Taylor-Couette flow) is particularly relevant.
Dynamo action is associated with spiral and sheared flows, which
suggests that Taylor-vortex flow is good candidate for dynamo
experiments in a simple geometry.

At the same time, Taylor-Couette flow is already a useful benchmark
for studying instabilities relevant to astrophysical processes
(Balbus \& Hawley 1991; Ji et al. 2001; R\"udiger \&  Zhang 2001),
as the flow pattern can model Keplerian motion.
In particular we note the works of \cite{dobler02} who studied the
dynamo mechanism in the presence of an imposed axial flow
(resulting in a screw dynamo) and of
\cite{laure00} who concentrated on the basic Taylor-Couette 
configuration.  The latter performed a kinematic dynamo calculation in
this geometry and demonstrated that an imposed Taylor-vortex flow is 
capable of dynamo action.
The combination of shear with simple roll flows has also been modelled
by \cite{dudley89} in spherical geometry.  There it was demonstrated
that magnetic field generation is sensitive to the nature of the 
driving flow.  Our work differs from the studies of \cite{dudley89} 
and \cite{laure00} in two important respects: first (kinematic dynamo) 
we use the velocity fields that are actual solutions of the
Navier-Stokes equations to generate a magnetic field, not arbitrary
imposed flow fields; secondly (fully self-consistent dynamos) we let
these velocity fields evolve alongside the magnetic field, thus 
solving the full MHD equations.

The aim of this work is two-fold.  Firstly we widen the 
study of \cite{laure00} in the parameter space
by finding the growth of the magnetic field as a function of 
the flow patterns.  Secondly, and more importantly,
we go beyond the limits of (linear) kinematic theory and investigate 
the fully self-consistent dynamo mechanism 
(magnetic fields and flow pattern affect each
other and saturate nonlinearly).

\section{Model}

We consider an incompressible fluid contained between two
coaxial cylinders of inner radius $R_1$ and outer radius $R_2$ which
rotate at prescribed angular velocities $\Omega_1$ and $\Omega_2$.
The velocity and magnetic fields,
$\vec{V}(r,\theta,z,t)$ and $\vec{B}(r,\theta,z,t)$
are determined by the MHD equations which we write in dimensionless
form as
\beq
   \label{eq:mom}
   \pder{\vec{V}}{t}{} + (\vec{V} \cdot \grad) \vec{V} =
   - \grad p + \laplace \vec{V} + (\curl \vec{B})
   \cross \vec{B},
\eeq \beq
   \label{eq:ind} 
   \pder{\vec{B}}{t}{} = \frac1{\Pm} \laplace \vec{B}
   + \curl (\vec{V} \cross \vec{B}),
\eeq \beq
   \dvgnce \vec{B}=0,\quad
   \dvgnce \vec{V}=0,
\eeq
where $p$ is the pressure.
In writing \rf{eq:mom} and \rf{eq:ind} we used
$\delta=R_2-R_1$ as unit of length, $\delta^2/\nu$ as unit of time
and $(\mu_0\rho)^\frac1{2}\nu/\delta$ as unit of  magnetic field, 
where the kinematic viscosity $\nu$, the density $\rho$ 
and the permeability $\mu_0$ are constant. 
The governing dimensionless parameters of the system
are the Reynolds numbers  $\Rey_1$, $\Rey_2$, the radius ratio $\eta$,
and rotation ratio $\mu$,
\beq
   \Rey_1 = \frac{R_1 \Omega_1 \delta}{\nu},\quad
   \Rey_2 = \frac{R_2 \Omega_2 \delta}{\nu},\quad
   \eta = \frac{R_1}{R_2},
   \quad \mu = \frac{\Omega_2}{\Omega_1},
\eeq
together with the magnetic Prandtl
number $\Pm$,
\beq
   \Pm=\frac{\nu}{\lambda},
\eeq
where $\lambda=1/(\mu_0\sigma )$ is the magnetic diffusivity and
$\sigma$ is the electrical conductivity.

We assume no-slip boundary conditions for $\vec{V}$, and electrically 
insulating boundary conditions for $\vec{B}$.  For modes of the form
$\vec{B}(r,t,z) = \vec{B}(r)\,\ex^{\im(\alpha z+m\theta)}$
these conditions are
\beq
   \label{eq:mag_bcs}
   \barr{rl}
   \alpha = m = 0:
   &
   B_\theta = B_z = 0;
   \\[2pt]
   \alpha=0, \, m\ne 0:
   &
   \pd{\theta} B_r = \pm m B_\theta,
   \quad
   B_z = 0 ;
   \\[2pt]
   \alpha \ne 0:
   &
   {\displaystyle
   \pd{z}B_r = \frac{\pd{r}\bess_m}{\bess_m} \, B_z ,
   \quad
   \frac{1}{r} \, \pd{\theta} B_z = \pd{z} B_\theta ,
   }
   \earr
\eeq
where for $\pm$ we take $+$ at $R_1$, $-$ at $R_2$.
The symbol $\bess_m(r)$ indicates the
modified Bessel functions,  
$I_m(\alpha r)$ at $R_1$ and $K_m(\alpha r)$ at $R_2$.
The axial wavelength of a pair of Taylor-vortices is $2\pi/\alpha$.

Our numerical method for timestepping the 3D nonlinear MHD equations 
is detailed in \cite{willis02}.  The formulation is
based on representing $\vec{V}$ and $\vec{B}$ with the toroidal-poloidal
decomposition
\beq
   \label{eq:pot_expansion}
   \vec{A} = \psi_0 \, \vechat{\theta} + \phi_0 \, \vechat{z}
   + \curl (\psi\vec{r}) + \curl \curl (\phi\vec{r}),
\eeq
where $\psi(r,t,z)$, $\phi(r,t,z)$ and $\psi_0(r)$, $\phi_0(r)$ 
contain the periodic and non-periodic parts of the field respectively.
The potentials are expanded spectrally over Fourier modes 
in the azimuthal and axial directions and over Chebyshev
polynomials in the radial direction.  

The governing equations for the magnetic field are the $r$-components
of the induction equation and its first curl.  For the velocity
we follow the procedure applied to the magnetic field and take the
$r$-components of the momentum equation and its first curl.  
As the pressure has not been eliminated, we also take the divergence
to obtain the pressure--Poisson equation. 
All five governing equations are second order in $r$.  
Time stepping is based on a
combination of second order accurate Crank-Nicolson and Adams-Bashforth
methods. The code has been tested against published
results with and without a magnetic field (Chandrasekhar 1961;
Roberts 1964; Marcus 1984; Jones 1985; Barenghi 1991).

\section{Kinematic dynamo}

For prescribed Reynolds number $\Rey_1$, wavenumber $\alpha$,
rotation ratio $\mu$ and radius ratio $\eta$ the 
Navier-Stokes Eq.\ \rf{eq:mom} is solved
in the absence of a magnetic field.  
We always assume $\Rey_1>\Rey_{1c}$, where $\Rey_{1c}$ is the critical
Reynolds number at which azimuthal Couette flow is unstable to
the formation of axisymmetric Taylor vortices (Chandrasekhar 1961).
We also assume $\alpha=3.14$, which corresponds to almost square cells.
The nonlinear Taylor-vortex flow $\vec{V}$ thus obtained is used when
solving the induction Eq. \rf{eq:ind} for $\vec{B}$, starting
from a small magnetic seed field of wavelength $\alpha_B$.

Note that $\vec{V}$ being fixed, \rf{eq:ind} is linear and has
eigenfunction solutions $\vec{B}$ which grow or decay exponentially.  If the
real part of the growth rate $\grB$ is positive then the magnetic
field grows (kinematic dynamo action).  
Following \cite{laure00}
we assume $\alpha_B=\frac1{2}\alpha$, the
characteristic length for the magnetic field is the length of
two \emph{pairs} of Taylor-vortices.  
In several calculations with $\alpha_B=\alpha$ we found
the magnetic seed field decayed quickly.
\begin{figure}
   \epsfig{figure=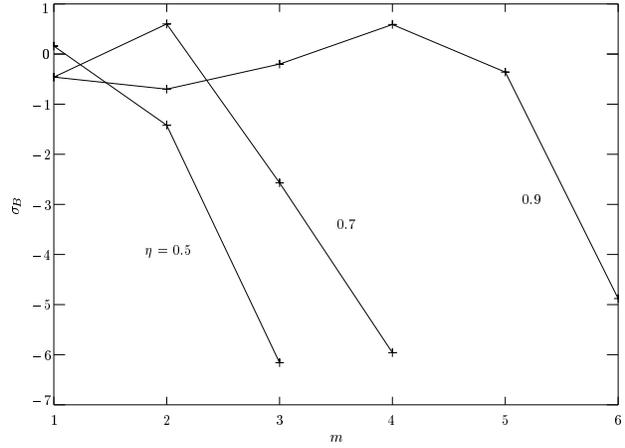, scale=0.56}
   \caption{\label{fig:grBetam} 
      Growth rates $\grB$ for various $\eta$ and $m$,
      where $\Rey_1=2\Rey_{1c}(\eta)$, $\Pm=2$.
   }
\end{figure}
The axisymmetric flow cannot generate an $m=0$ magnetic 
field.  
In Fig.\ \ref{fig:grBetam} we see that in narrower gaps the dynamo
prefers larger $m$.
The dynamo is local here in the sense that the characteristic length
scale for the magnetic field comply with the scale of the flow.
As the Taylor-vortex flow is itself unstable to non-axisymmetric
perturbations in narrower gaps (see Jones 1985), we usually consider
the case $\eta=0.5$ where $m=1$ is preferred.  The relative stability of
Taylor-vortex flow in wider gaps also allows for a clearer 
interpretation of the affect of the magnetic field in the nonlinear
self-consistent solutions presented in Sect.\ \ref{sect:scdyn}.

\begin{figure}
   \epsfig{figure=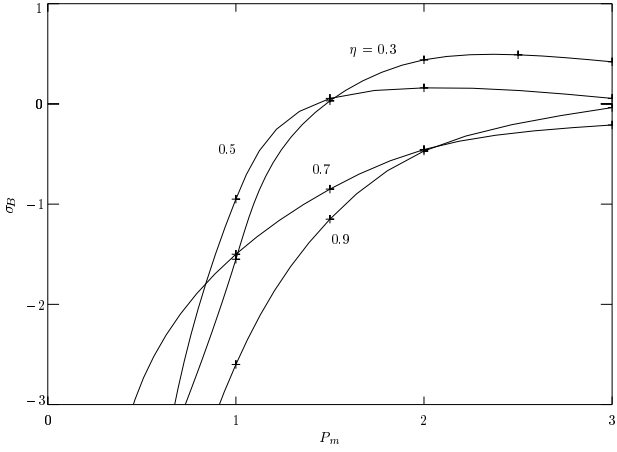, scale=0.56}
   \caption{\label{fig:grBpreta} 
      Growth rates $\grB$ for various $\eta$ as a function of $\Pm$,
      where $\Rey_1=2\Rey_{1c}(\eta)$, $m=1$.  Critical Reynolds numbers
      at $\eta=0.3, 0.5, 0.7, 0.9$ are respectively
      $\Rey_{1c} = 73.3, 68.2, 79.5, 131.6$.
   }
\end{figure}
Figure \ref{fig:grBpreta}
shows that, not surprisingly, $\grB$ falls off sharply when $\Pm<1$,
in which case much larger Reynolds numbers (relative to $\Rey_{1c}$)
are needed for dynamo action.
Generally it is easier to generate a magnetic field in media with 
larger magnetic Prandtl numbers.  
This is believed to be the case for warm weakly ionised gas, 
\cite{mcivor77}, \cite{kulsrud92}.

\begin{figure}
   \epsfig{figure=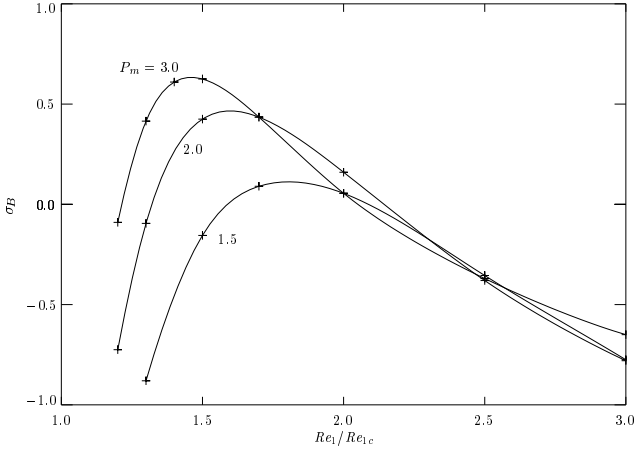, scale=0.56}
   \caption{\label{fig:grBRe1pr}
      Growth rates $\grB$ as versus $\Rey_1$ at $\eta=0.5$.
   }
\end{figure}
\begin{figure}
   \epsfig{figure=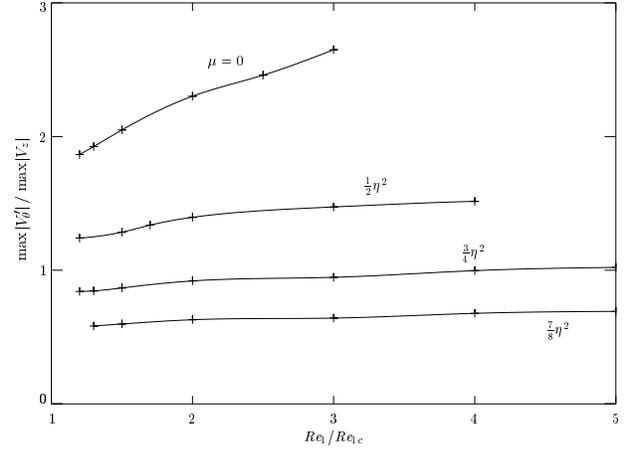, scale=0.56}
   \caption{\label{fig:utuzRe1co}
      Amplitude of the azimuthal disturbance 
      $V'_\theta=V_\theta-r\,\Omega(r)$ drops as $\mu\to\eta^2$ in 
      the absence of a magnetic field.
      At each $\Rey_1/\Rey_{1c}$ the amplitudes of $V_z$ are close 
      for all $\mu$.
   }
\end{figure}

Figure \ref{fig:grBRe1pr}
shows growth rates as a function of $\Rey_1$ with $\Rey_2=0$, $\eta=0.5$
for a few values of $\Pm$.  Driving the flow harder seems to move the
flow into a regime which is less favourable for magnetic field generation
when the outer cylinder is fixed.
Allowing co-rotation, the stability boundary for onset of Taylor-vortex
flow tends to the Rayleigh line $\mu=\eta^2$ for large Reynolds numbers.
As circular-Couette flow $\Omega(r)=a+b/r^2$ alone is not capable of 
dynamo action, we must consider flows which are Rayleigh unstable,
 $\mu<\eta^2$.
\begin{figure}
   \epsfig{figure=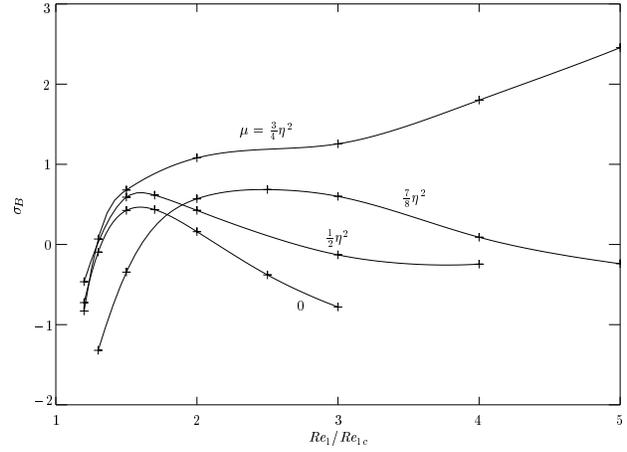, scale=0.56}
   \caption{\label{fig:grBRe1mu}
      Growth rates $\grB$ with co-rotation.  $\eta=0.5$, $\Pm=2$.
      At $\mu=0,\, \frac1{2}\eta^2,\, \frac{3}{4}\eta^2,\, 
      \frac{7}{8}\eta^2$, critical Reynolds numbers are 
      $\Rey_{1c} = 68.2, 84.3, 112.7, 155.3$ respectively.
   }
\end{figure}
Figure \ref{fig:grBRe1mu}
shows growth rates as a function of $\Rey_1$ for 
different values of $\mu$.  Again, we normalise the 
intensity of the drive $\Rey_1$ by the critical value $\Rey_{1c}$.
It is apparent that magnetic field generation is easier with 
co-rotating cylinders 
and $\mu=\frac{3}{4}\eta^2$ is an important case.
At this particular $\mu$ the amplitude of the azimuthal disturbance
$V'_\theta=V_\theta-r\,\Omega(r)$ and $V_z$ are similar, 
seen in Fig.\ \ref{fig:utuzRe1co}.
At a given $\Rey_1/\Rey_{1c}(\mu)$,  the components $V_r$, $V_z$
appear to be very similar in structure and amplitude.  
However, $V_\theta$ changes significantly with $\mu$.

\begin{figure}
   \epsfig{figure=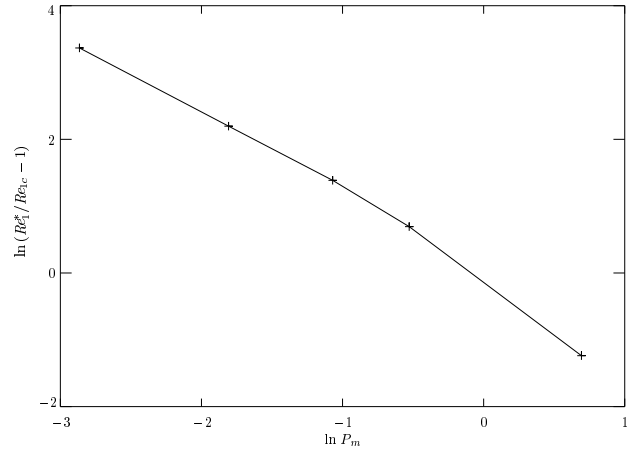, scale=0.56}
   \caption{\label{fig:Re1bpr}
      Dependence on $\Pm$ for the driving required for field
      generation. $\mu=\frac{3}{4}\eta^2$, $\eta=0.5$.
   }
\end{figure}
Of the rotation ratios considered in Fig.\ \ref{fig:grBRe1mu},
$\mu=\frac{3}{4}\eta^2$ 
would appear to be most favourable for
magnetic field generation.  
We define $\Rey_1^*$
as the critical Reynolds number at which $\grB=0$ 
(marginal state for the onset of dynamo action) for this $\mu$.
Figure \ref{fig:Re1bpr}
shows how $\Rey_1^*$ depends on $\Pm$.
Fitting the last three points for small $\Pm$ we obtain the slope
$-1.1$.
Approximately, $\Rey_1^*/\Rey_{1c}\propto 1/ \Pm$, for small $\Pm$.
Our interest in this case is linked to the small magnetic Prandtl
numbers of laboratory fluids.
Defining the magnetic Reynolds number $\Rm$ as
\beq
   \Rm = \frac{R_1\,\Omega_1\,\delta}{\lambda} = \Rey_1 \,\Pm
\eeq
we conclude that the critical magnetic Reynolds number 
$\Rm^*=\Rey_1^*\, \Pm$ is approximately constant and 
${\mathcal{O}}(10^2)$
for suitably chosen $\eta$, $\mu$.
Here $\mu\approx\frac{3}{4}\eta^2$ 
but the most suitable $\mu$ is likely to vary for other $\eta$.
The $\Rm^*$ above are also consistent with the results of \cite{laure00} 
who found $\Rm^* = {\mathcal{O}}(10^2)$ in their calculations 
with $\mu=0$.

\section{Self-consistent dynamo}
\label{sect:scdyn}

In this section Eqs.\ \rf{eq:mom} and \rf{eq:ind} are solved
simultaneously using as initial conditions the previously prescribed
Taylor-vortex flow $\vec{V}$ and the accompanying eigenfunction $\vec{B}$.
\begin{figure}
   \epsfig{figure=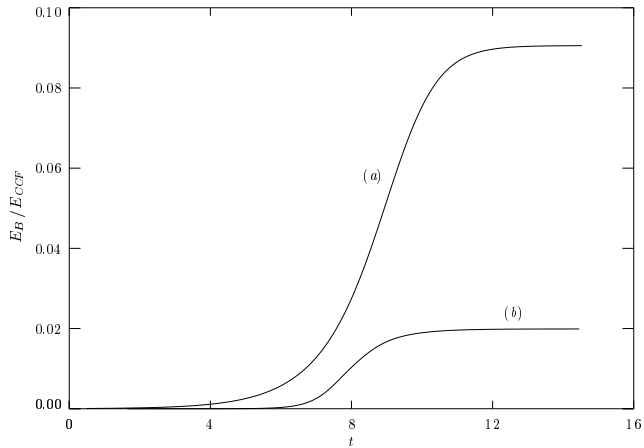, scale=0.56}
   \caption{\label{fig:benergy}
      Magnetic energy versus time as the magnetic field saturates. 
      $\eta=0.5$, $\Pm=2$. 
      ({\it a}) 
      $\mu=0$, $\Rey_1=1.5\Rey_{1c}$, 
      $\Eccf = 4.15\times 10^4$.
      ({\it b})       
      $\mu=\frac{3}{4}\eta^2$, $\Rey_1=2\Rey_{1c}$,
      $\Eccf = 3.63\times 10^5$.
   }
\end{figure}
Figure \ref{fig:benergy} 
shows that after an initial transient the magnetic energy $E_B$
saturates to 
a constant value, indicating dynamo action.  
At the parameters in Fig.\ \ref{fig:benergy} 
the calculations required 16 Chebyshev modes radially, 
16 axial and 12 azimuthal Fourier modes with a timestep of $10^{-4}\tau$,
where $\tau$ is the rotation period of the inner cylinder.
The magnetic energy is plotted as a fraction of the energy of 
the driving circular-Couette flow $\Eccf$,
the energy source for the dynamo.  The energy sink is an
increased viscous dissipation in the velocity disturbance
in addition to Ohmic dissipation.
In these calculations the dynamo is
dynamically self consistent ($\vec{V}$ and $\vec{B}$ evolve together).

\begin{figure}
   \bcent
      \epsfig{figure=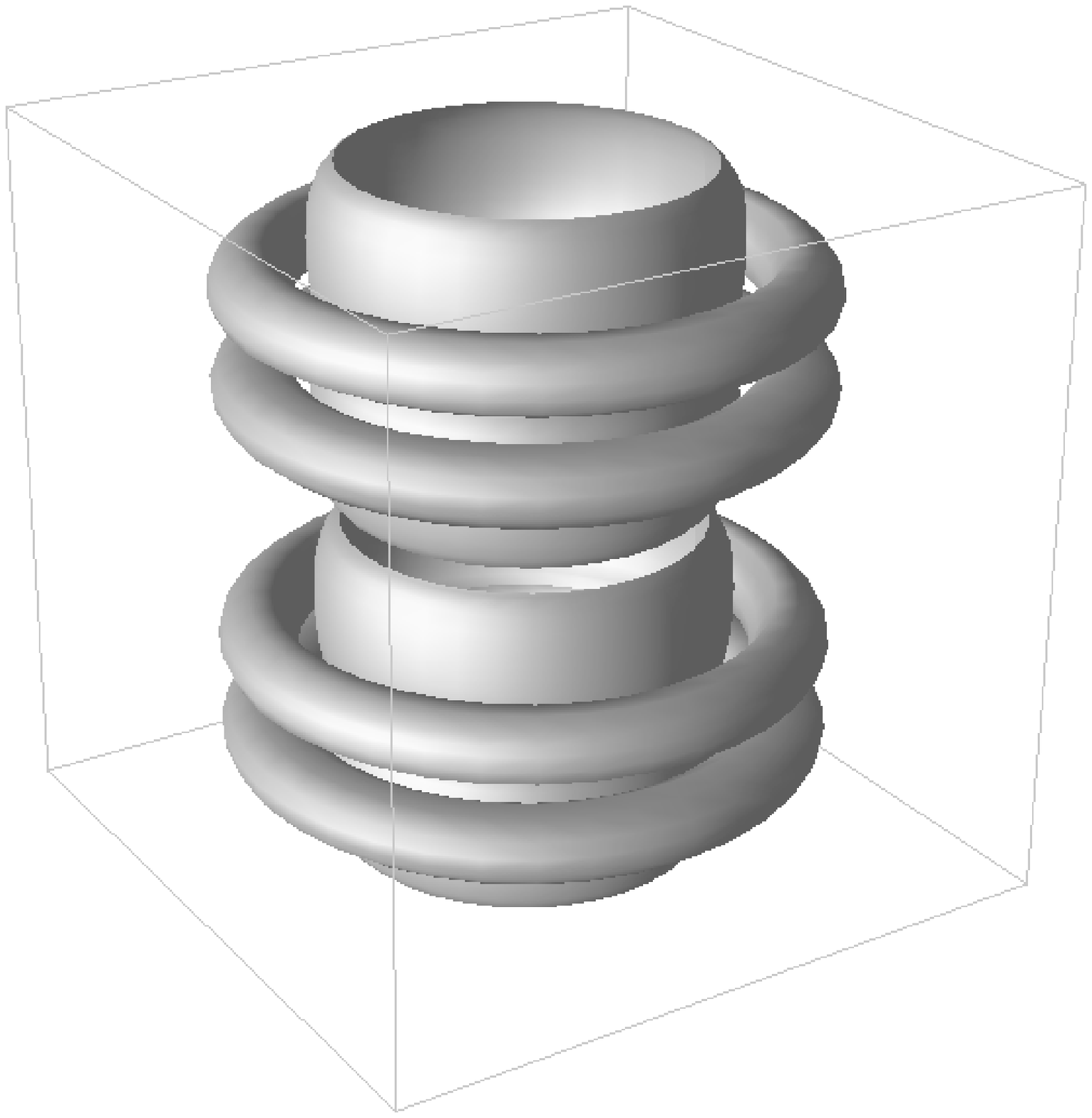, scale=0.35, angle=270}
   \ecent
   \caption{\label{fig:eighel} 
      Isosurface of helicity $|\vec{V}\cdot\curl\vec{V}|$ at $\eta=0.5$,
      $\Rey_1=1.5\Rey_{1c}$, $\mu=0$.  Shown over two axial
      periods.
   }
\end{figure}

\begin{figure}
   \bcent
      \epsfig{figure=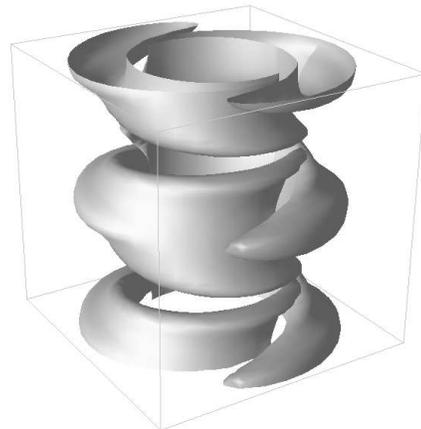, scale=0.35, angle=270}
   \ecent
   \caption{\label{fig:eigB} 
      Isosurface of $|\vec{B}|$.  The field is $m=1$.  Same parameters as
      Fig.\ \ref{fig:eighel} with $\Pm=2$ 
      (as in Fig.\ \ref{fig:benergy}{\it a}).
   }
\end{figure}

\begin{figure}
   \bcent
      \epsfig{figure=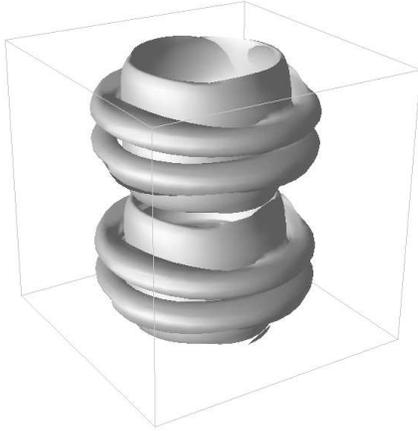, scale=0.35, angle=270}
   \ecent
   \caption{\label{fig:sathel} 
      Isosurface of helicity at magnetic field saturation.  
      The flow is $m=2$, looking the same if rotated by $180^\circ$.
      Parameters as in Fig.\ \ref{fig:benergy}{\it a}.
   }
\end{figure}

Figures \ref{fig:eighel} and \ref{fig:eigB}
show the typical field structure of the initial conditions.
A fixed outer cylinder was taken for visualisation purposes
as the surfaces are less self-obscuring.  The initial
flow pattern $\vec{V}$ has $m=0$ symmetry and as $\Rey_1>\Rey_{1c}$
vortex cores are slightly shifted towards the outflow regions.
The eigenfunction $\vec{B}$ in Fig.\ \ref{fig:eigB}
has $m=1$ symmetry.  The flow pattern, initially axisymmetric is
deformed by the action of the Lorentz force $(\curl\vec{B})\cross\vec{B}$
and acquires an $m=2$ contribution visible in Fig.\ \ref{fig:sathel}.
\begin{figure}
   \epsfig{figure=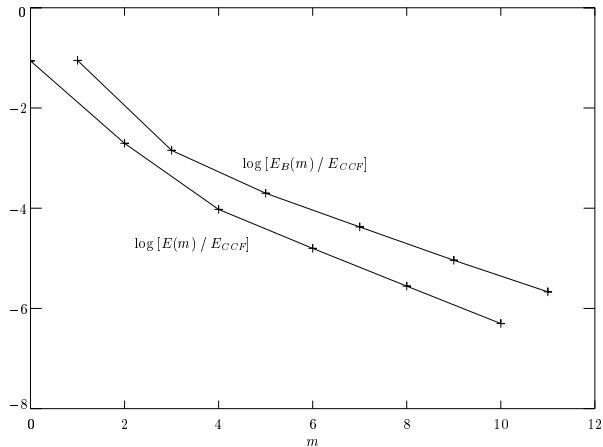, scale=0.56}
   \caption{\label{fig:em}
      Kinetic energy of the velocity disturbance and the magnetic energy 
      of the various azimuthal modes.  Parameters as in 
      Fig.\ \ref{fig:benergy}{\it a}.
   }
\end{figure}
Figure \ref{fig:em} shows the kinetic energy of the velocity disturbance
and the magnetic energy of the various azimuthal modes, 
$E(m)$ and $E_B(m)$ respectively, for the saturated
dynamo of Fig.\ \ref{fig:benergy}{\it a}.  The velocity has contributions
$m=0,2,4,\dots$ and the magnetic field has $m=1,3,5,\dots$ etc.
The perturbation to the magnetic field is difficult 
to appreciate visually on the dominant $m=1$ structure.  
It remains rather similar to Fig.\ \ref{fig:eigB}.

\section{Discussion}
By solving the kinematic dynamo problem, we have determined that a
Taylor-vortex flow pattern can sustain a growing magnetic field.

Like in the models of \cite{dudley89} we also find that the dynamo
is sensitive to the flow pattern.  Further, for flows that are capable
dynamo action we see that the growth rate
is not a monotonic increasing function of the Reynolds number.  This
is not seen in \cite{dudley89}, most likely due to the prescribed form
for the driving flow patterns.

In the Taylor-vortex flow the best growth rates have been obtained 
with co-rotation.
The relative magnitude of the shear and roll in the flow plays an 
important part in the success of the dynamo mechanism.

Solving the full MHD 
equations we have demonstrated the existence of a fully self-consistent 
nonlinearly saturated dynamo.  
Hopefully these results will stimulate experimental
work on the problem.  
Future theoretical work will investigate dynamo action in 
hydrodynamically stable flows and address the
nature of the magnetic field structure when the dynamo is driven
harder -- our dynamo is laminar.  
Most of the present work is concerned with wider gaps.


\end{document}